\documentclass[10pt]{iopart}
\usepackage{epsf}
\def\a{\alpha}
\def\b{\beta}
\def\g{\gamma}
\def\G{\Gamma}

\def\D{\Delta}

\def\w{\omega}

\begin{document}

\title[]{Quantum theory of shuttling instability in a movable quantum dot array}

\author{Andrea Donarini\dag,\ Tom\'a\v s Novotn\'y\dag,\ddag\ and Antti--Pekka Jauho\dag}

\address{\dag\ Mikroelektronik Centret,
         Technical University of Denmark,
         \O rsteds Plads, Bldg.\ 345 east,
         DK-2800 Kgs.\ Lyngby, Denmark}

\address{\ddag\ Department of Electronic Structures,
     Faculty of Mathematics and Physics, Charles University,
     Ke Karlovu 5, 121 16 Prague, Czech Republic}

\eads{\mailto{ad@mic.dtu.dk},\mailto{tno@mic.dtu.dk}}

\begin{abstract}
We study the shuttling instability in an array of three quantum
dots the central one of which is movable. We extend the results by
Armour and MacKinnon on this problem to a broader parameter
regime. The results obtained by an efficient numerical method are
interpreted directly using the Wigner distributions. We emphasize
that the instability should be viewed as a crossover phenomenon
rather than a clear-cut transition.
\end{abstract}



%\maketitle

\section{Introduction}
Since the first proposal by Gorelik et al.\ \cite{gor-prl-98} of
the shuttling instability in a generic nanoelectromechanical
system (NEMS) consisting in a movable single electron transistor
the phenomenon of the shuttling transport attracted much
attention. However, until recently the fully quantum theory of the
phenomenon was not developed. The first quantum mechanical study
on a modified setup was accomplished by Armour and MacKinnon
\cite{arm-prb-02}, closely followed by the work of the present
authors on the original Gorelik's setup \cite{PRL}. In this paper
we extend the results by Armour and MacKinnon. The phase space
analysis in terms of the Wigner functions introduced in \cite{PRL}
reveals directly the nature of the transport (incoherent
tunnelling versus shuttling) in different regions in contrast to
the indirect evidence used by Armour and MacKinnon. Also the new
numerical scheme that we use enables us to access a wider range of
parameters fast and reliably.

\section{Model and method of solution}
Let us consider a simple NEMS consisting of an array of three
quantum dots (device) connected to two leads \cite{arm-prb-02}.
The central dot is assumed to be movable in a parabolic potential.
The Hamiltonian of the device consists of the mechanical and the
electronic parts $H=H_{\rm mech} + H_{\rm el}$ where $H_{\rm mech}
=\hbar \w \hat{a}^{\dagger} \hat{a}$ is the Hamiltonian of the
harmonic oscillator and $H_{\rm el}=\sum_{\a,\b\in \{0,L,C,R\}}|\a
\rangle \epsilon_{\a\b}(\hat{x})\langle \b|$. We assume strong
Coulomb blockade regime with no double occupancies so that the
vectors $|\a\rangle$ with $\a=0,L,C,R$ span the entire electronic
Hilbert space of the device. Each matrix element
$\epsilon_{\a\b}(\hat{x})$ is still a full matrix in the
oscillator space. $\epsilon(\hat{x})$ reads explicitly
\begin{equation}
 \epsilon(\hat{x})=\left[\begin{array}{cccc}
   0 & 0 & 0 & 0 \\
   0 &\frac{\D V}{2} & t_L(\hat{x}) & 0 \\
   0 & t_L(\hat{x}) & -\frac{\D V}{2x_0}\hat{x} & t_R(\hat{x}) \\
   0 & 0 & t_R(\hat{x}) & -\frac{\D V}{2}
\end{array}\right]
\end{equation}
where
$\hat{x}=\sqrt{\frac{\hbar}{2m\w}}(\hat{a}+\hat{a}^{\dagger})$ is
the position operator, $\D V$, the {\it device bias}, is the
difference between the energy of the left and the right dot. $x_0$
is half the distance between the two outer dots and represents the
maximum amplitude of the inner dot oscillation. The three dots are
electrically connected only via a tunnelling mechanism. The
tunnelling length is given by $1/\a$ and the tunnelling strengths
depend on the position operator $\hat{x}$ of the inner grain as
$t_L(\hat{x})=V_0e^{-\a(x_0+\hat{x})},\,
t_R(\hat{x})=V_0e^{-\a(x_0-\hat{x})}$.

The dynamics of the device is described by the generalized master
equation (GME) formalism \cite{arm-prb-02}. The density matrix of
the device evolves according to $\dot{\rho}=-i[H,\rho]+ \Xi \rho +
\dot{\rho}_d$. The first term represents the coherent evolution of
the isolated device. The coupling to the leads responsible for the
electronic transfer from/to the device is introduced in the wide
band approximation following Gurvitz et al. \cite{gur-prb-96} and
yields the second term in the equation (in the block notation of
\cite{arm-prb-02})
\begin{equation}
 \Xi \rho=\G\left[\begin{array}{cccc}
 \rho_{RR}-\rho_{00} & 0            & 0             & 0 \\
   0                 &    \rho_{00} & 0             & -\rho_{LR}/2 \\
   0                 & 0            & 0             & -\rho_{CR}/2 \\
   0                 & -\rho_{RL}/2 & -\rho_{RC}/2  & -\rho_{RR}   \\
 \end{array}\right]
\end{equation}
with $\G$ being the injection rate to the leads. The third term
describes the effect of the environment on the oscillator,
consisting in mechanical damping and random quantum and thermal
excitation (Langevin force). It reads \cite{arm-prb-02}
\begin{equation}\fl
 \dot{\rho}_d = -\frac{\g}{2}\bar{n}(a a^{\dagger}\rho - 2 a^{\dagger}\rho a + \rho a a^{\dagger})
                -\frac{\g}{2}(\bar{n}+1)( a^{\dagger}a\rho - 2 a \rho a^{\dagger} + \rho a^{\dagger}a)
\end{equation}
where $\gamma$ is the damping rate and $\bar{n}$ is the mean
occupation number of the oscillator at temperature $T$.

The stationary version of the above GME was solved numerically
after the oscillator Hilbert space was truncated at sufficiently
large $N$ so that all dynamically excited states were contained
within the basis. Utilizing the decoupling properties of the GME
in the block notation resulted in the problem of finding a unique
null space of a (super)matrix with the linear dimension $10N^2$
(with up to $N=40$). The Arnoldi iteration \cite{golub} with
preconditioning was found to be superior to direct methods, such
as singular value decomposition or inverse iteration. The
preconditioning involving the inversion of the Sylvester part of
the problem (a fast procedure) is crucial for the convergence of
the method which is very fast and low memory consuming (the whole
(super)matrix does not have to be stored in the memory).

\begin{figure}
\begin{center}
\epsfxsize=\textwidth \epsfbox{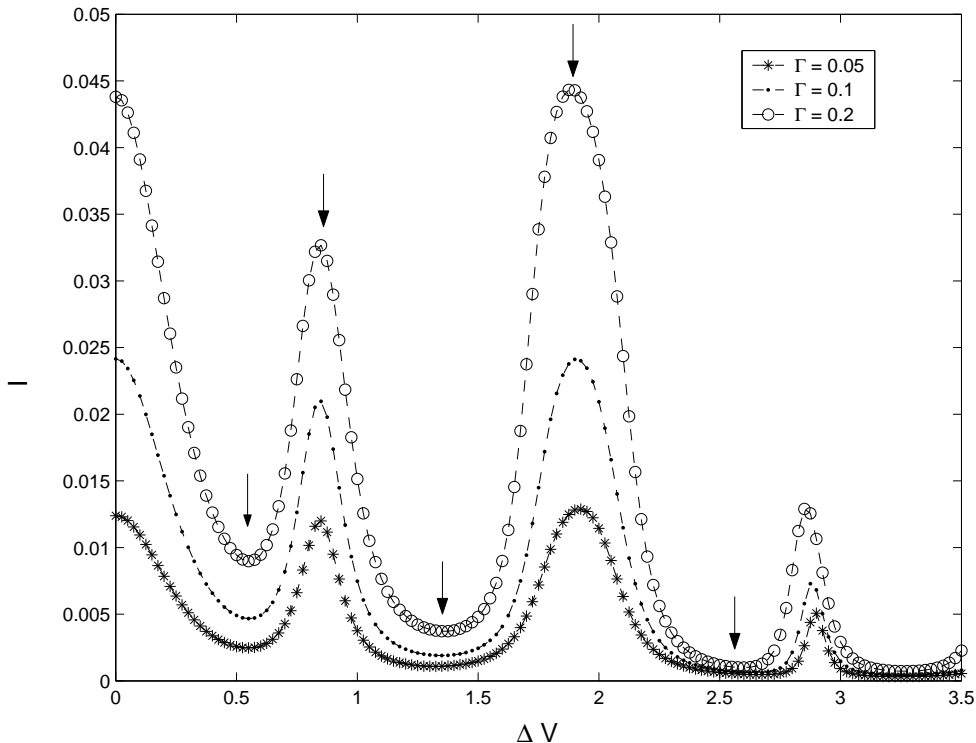}
\end{center}
\caption{Plot of the current through the triple dot system as a
function of the device bias for different injection rates
$\Gamma$. The other parameters are $V_0 = 0.5$, $\a = 0.2 $,
$x_0=5$, $\g=0.0125$, $\w=1$.}\label{equilsol}
\end{figure}

\section{Results and discussion}
The resulting stationary density matrix was used to evaluate the
mean value of current $I = \G\,{\rm Tr}_{\rm osc}\rho_{RR}$
\cite{arm-prb-02} and the phase space distribution of the charged
central dot using the transformation into Wigner coordinates
\cite{PRL}. These quantities as a function of the device bias $\D
V$ and for three different values of the injection rate $\G$ are
plotted in the figures 1 and 2, respectively. It was found in
\cite{arm-prb-02} that the triple dot system exhibits different
regimes of transport at different device biases. The current peaks
at $\D V \approx n \w$ (see figure 1) were identified as effects
of electromechanical resonances within the device. Yet, the
different peaks may correspond to different physical mechanisms
--- while the peak around $\D V \approx \w$ is mainly due to the
incoherent oscillator-assisted tunnelling the peak at $\D V
\approx 2\w$ reveals a clear shuttling component. This finding by
Armour and MacKinnon based on indirect evidence of parametric
dependencies of the current curves (e.g.\ the dependence of the
current curve on the tunneling length $1/\a$) is confirmed by the
direct inspection of the phase space distributions (see the first
row of figure 2). The half-moon-like shape characteristic for
shuttling transport \cite{PRL} is only present around $\D V
\approx 2\w$ while all other plots show just the fuzzy spot
indicative of incoherent tunnelling. However, our direct criterion
for detecting the shuttling regime reveals a close similarity
between the resonances. For increasing injection rate we can see
that the shuttling regime gradually sets in also in the vicinity
of the first resonance peak. This reveals the crossover character
of the onset of the shuttling instability found also previously
\cite{PRL}. The sharp transition into the shuttling regime
reported in semiclassical studies is smeared into the crossover
due to the noise present in the system and properly accounted for
in our approach.

The authors thank T.~Eirola for indispensable advice concerning
the numerical methods. Support of the grant 202/01/D099 of the
Czech grant agency for one of us (T.N.) is also gratefully
acknowledged.

\begin{figure}
\begin{center}
\epsfxsize=\textwidth \epsfbox{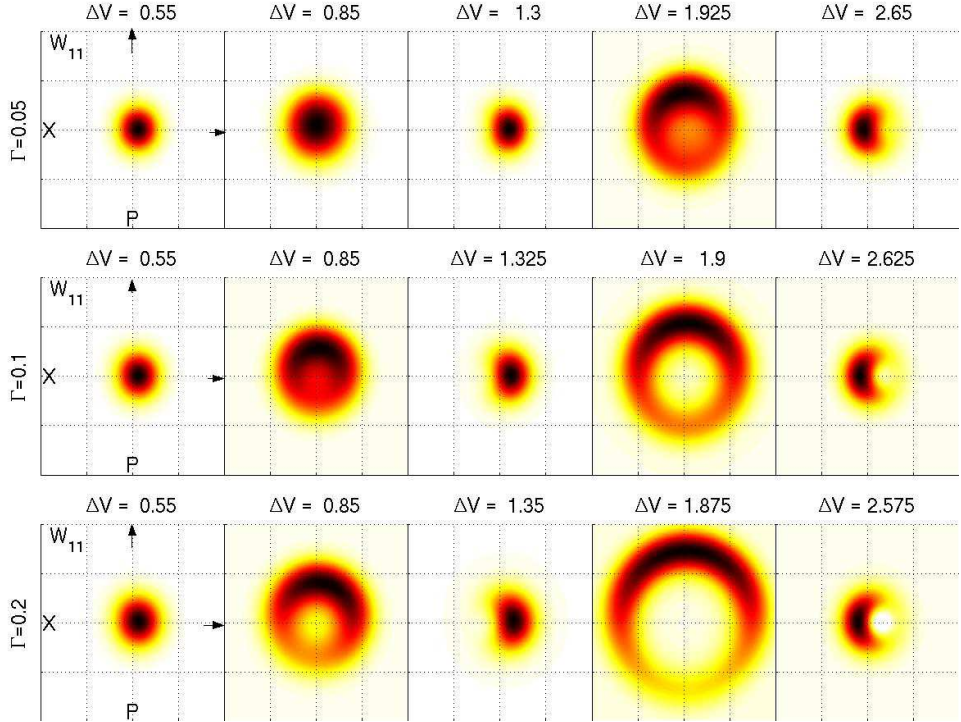}
\end{center}
\caption{Wigner distributions for the central dot in the charged
state. Different device biases are the points of minimum or
maximum current (see figure 1).}\label{wigner}
\end{figure}

%\subsection{Acknowledgments}

%\bibliographystyle{unsrt}
%\bibliography{shuttling}

\end{document}